\documentclass[12pt]{article}

\usepackage{epsfig} 
\usepackage{epstopdf}
\usepackage{amssymb}
\usepackage{amsmath}
\usepackage{amsfonts}
\usepackage{graphicx}
\graphicspath{ {plots/} }
\usepackage{mathrsfs}
\usepackage{color}
\usepackage{multirow}

\newcommand{\fu}{\mathfrak{u}}

\newcommand{\fn}{{\mathfrak{n}}}

\newcommand{\fz}{\mathfrak{z}}

\newcommand{\bM}{\mathbf{M}}

\newcommand{\cE}{\mathcal{E}}

\newcommand{\cJ}{\mathcal{J}}

\newcommand{\cP}{\mathcal{P}}

\newcommand{\cT}{\mathcal{T}}

\newcommand{\be}{\begin{equation}}
\newcommand{\ee}{\end{equation}}
\newcommand{\bea}{\begin{eqnarray}}
\newcommand{\eea}{\end{eqnarray}}
\newcommand{\nn}{\nonumber}

\newcommand{\ed}{\end{document}}

\newcommand{\bi}{\begin{itemize}}
\newcommand{\ei}{\end{itemize}}

\newcommand{\bce}{\begin{center}}
\newcommand{\ece}{\end{center}}

\begin{document}

\title{Broadband Coherent Perfect Absorber with $\mathcal{P}\mathcal{T}$-Symmetric 2D-Materials}

\author{Mustafa Sar{\i}saman$^{1,}$\thanks{Email Address: mustafa.sarisaman@istanbul.edu.tr} and Murat~Tas$^2$\thanks{Email Address: tasm236@gmail.com}\\[6pt]
$^1$Department of Physics, Istanbul University, 34134 Istanbul, Turkey\\
$^2$Department of Software Engineering, Altinbas University, 34217 Istanbul, Turkey}

\date{ }
\maketitle

\begin{abstract}

We suggest graphene and a two-dimensional (2D) Weyl semimetal (WSM) as 2D materials for the realization
of a broadband coherent perfect absorber (CPA) respecting overall $\mathcal{PT}$-symmetry. We also
demonstrate the conditions for mutually equal amplitudes and phases of the left and right incoming waves
to realize a CPA. 2D materials in our system play the role to enhance the absorption rate of a CPA once
the appropriate parameters are inserted in the system. We show that a 2D WSM is more effective than
graphene in obtaining the optimal conditions. We display the behavior of each parameter governing the
optical system and show that optimal conditions of these parameters give rise to enhancement and
possible experimental realization of a broadband CPA-laser.
\vspace{2mm}



\end{abstract}

\section{Introduction}

Time-reversed consideration of regular lasers come into existence by purely ingoing fields, which is
known as coherent perfect absorption (CPA), or antilaser~\cite{antilaser1,antilaser2,antilaser2-1,
antilaser2-2,antilaser2-3,antilaser3,antilaser4,antilaser5}. This intriguing phenomenon has made a
rather drastic influence in nanostructured optical materials and especially in  plasmonics~\cite{cpa1},
and recently has attracted a broad interest. They underlie many applications, including molecular
sensing, photocurrent generation and photodetection~\cite{antilaser5}. In a CPA, complete absorption at
a single frequency can be achieved by illuminating two counter propagating fields~\cite{cpa2,lastpaper},
see~\cite{antilaser5} for a recent review of CPAs. Time reversal symmetry reveals the condition that a
CPA supports the self-dual spectral singularities~\cite{jpa-2012,naimark,naimark-1,naimark-2}. Since
the time reversal symmetry incorporates CPA action to lasing threshold condition, it comes out to be a
natural consequence of $\mathcal{PT}$-symmetric potentials that support both CPA and laser actions
simultaneously~\cite{antilaser2-1}. This makes $\mathcal{PT}$-symmetric CPA-lasers one of the primary
examples in the study of optics~\cite{jpa-2012}, and also rather intriguing because of its function as
a laser emitting coherent waves unless it is subject to incident coherent waves with appropriate
amplitude and phase in which case it acts as an absorber~\cite{lastpaper}. In this work, we investigate
the prospect of realizing a broadband CPA-laser in a linear homogeneous $\mathcal{PT}$-symmetric
optical system covered by two-dimensional (2D) materials. We set out the prescribed 2D materials as the
absorbing medium, and employ their prominent features in smooth experimental achievement of a CPA which
consists of equal amplitudes and phases of incoming waves.

$\mathcal{PT}$-symmetry in optical systems is achieved by means of complex refractive indices, such that
their optical modulations in complex dielectric permittivity plane results in both optical absorption
and amplification. $\mathcal{PT}$-symmetry is rather practical in optics since it helps related
parameters of the optical system be adjusted properly. Thus, $\mathcal{PT}$-symmetric optical
systems~\cite{bender,bender-1,bender-2} are currently studied actively due to their applications in a
series of intriguing optical phenomena and devices, such as dynamic power oscillations of light
propagation, lasers~\cite{naimark,naimark-1,naimark-2,p123,longhi4}, CPA lasers~\cite{lastpaper,CPA,
CPA-1,CPA-2,CPA-3,pra-2015d,yan}, and unidirectional invisibility~\cite{bender,bender-1,bender-2,PT6,
PT7,pra-2017a,paper1}. CPA phenomenon has become one of leading applications of
$\mathcal{PT}$-symmetric optical potentials since its emergence~\cite{antilaser1}.

\begin{figure}
    \begin{center}
    \includegraphics[scale=.60]{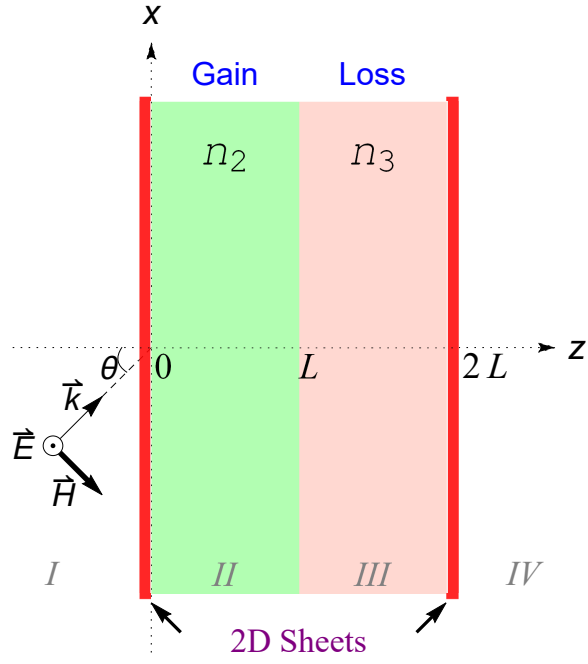}
    \caption{(Color online) TE mode configuration for the parallel pair of optically active system
covered by the 2D material sheets obeying the $\mathcal{PT}$-symmetry.}
    \label{fig1}
    \end{center}
    \end{figure}

Discovery of graphene led to over a decade of its intense study. It has triggered development of a vast
area of research on a variety of 2D materials, whose properties substantially differ from those of their
bulk counterparts~\cite{2d1}. Graphene is arguably the most famous 2D material of the last decade, and
fascination with its properties has spread beyond the scientific community. Emergence of graphene has
led to arise a voluminous literature and numerous applications has been carried out in various
fields~\cite{gr5,gr5-1,gr5-2,gr6,gr7,gr8,gr9,naserpour}. As the family of 2D materials expanded to
include new members such as 2D Weyl semimetals (WSM), 2D semiconductors, boron nitride and more
recently, transition metal dichalcogenides and Xenes, atomically thin forms of these materials offer
endless possibilities for fundamental research, as well as demonstration of improved or even entirely
novel technologies~\cite{2d2,2d2-1,2d2-2}. In view of these exciting properties of 2D materials
together with the idea that they may interact with electromagnetic waves in anomalous and exotic ways,
providing new phenomena and applications, the new distinctive studies of laser and CPA phenomena with
2D materials have arisen. Especially recent works on this field fashion up essential motivation of our
work~\cite{lastpaper, graphene, graphene-1, graphene-2, graphene-3, graphene-4}, which will use the
whole competency of transfer matrix method in a scattering formalism~\cite{prl-2009}. In this study,
we offer one of the potential applications of 2D materials in the field of CPA laser actions together
with fascinating $\mathcal{PT}$-symmetry attribute in optical systems. Our system is depicted in
Fig.~\ref{fig1}.

We employ a one-dimensional $\mathcal{PT}$-symmetric optically active system which is covered by a 2D
material, which respects the entire $\mathcal{PT}$-symmetry, see Fig.~\ref{fig1}. Since CPA conditions
coincide with the lasing threshold conditions, both of which could be expressed by means of spectral
singularities, the former is typically expressed by self-dual spectral singularities which is complex
conjugate of spectral singularities. We look for practical ways to improve the efficiency of CPA
through the supplemented coating materials. Although our formalism is satisfied by all 2D materials,
for the likeliest demonstration we use graphene and 2D WSM, and compare their impacts saliently.

We reveal complete solutions, schematically demonstrate their behaviors and show the effects of various
parameter choices yielding CPA conditions. We demonstrate that optimal control of parameters of the slab
(gain coefficient, incidence angle and slab thickness). Relevant parameters of 2D material give rise to
a desired outcome of achieving enhancement of broadband absorption, and computing correct amplitude and
phase contrasts in a CPA-laser. We present exact conditions causing the achievement of CPA with equal
amplitude and phase values of ingoing waves. We display the roles of 2D material in the gain decrement,
broadband absorption and reciprocities of amplitudes and phases in the production a CPA action. Results
of this study show that 2D WSM is more appropriate compared to graphene since it leads to build up a
CPA with minimum gain values at a critical $b$ value, below which no effect is observed. Also, 2D WSM
helps to improve the broadband accessibility of CPA based on the corresponding parameter adjustments.
We provide certain parameters belonging to our system setup if one desires experimental realization of
a broadband CPA with equal wave amplitudes and phases.

\section{TE Mode Solution, Transfer Matrix and CPA Condition}\label{S2}

TE mode Helmholtz equation describing our optical setup in Fig.~\ref{fig1} reads as
    \begin{align}
    &\left[\nabla^{2} +k^2\fz_{j}(z)\right] \vec{E}^{j}(\vec{r}) = 0,
    &&\vec{H}^{j}(\vec{r}) = -\frac{i}{k Z_{0}} \vec{\nabla} \times \vec{E}^{j}(\vec{r}).
    \label{equation4}
    \end{align}
Its solution yields TE waves in the form
    \begin{align}
    \vec E^{j}(\vec{r})=\mathcal{E}^{j} (z)e^{i k_x x}\hat e_y.
    \label{ez1}
    \end{align}
In this formulation, $\vec r:=(x,y,z)$ represents the coordinates, $k:=\omega/c$ is the wavenumber,
$c:=1/\sqrt{\mu_{0}\varepsilon_{0}}$ is the speed of light in vacuum,
$Z_{0}:=\sqrt{\mu_{0}/\varepsilon_{0}}$ is the impedance of the vacuum, $\hat e_y$ is the unit vector
in $y$-direction, $k_x$ and $k_z$ are the components of wavevector $\vec{k}$ in $x-z$ plane. Complex
quantity $\fz_{j}(z)$ is denoted by
    \begin{align}
    \fz_{j}(z):= \fn_j^{2} ~~~~~ {\rm for~}z\in z_{j}.
    \label{e1}
    \end{align}
The index $j$ represents the regions in Fig.~1 and $j = 1, 2, 3,~\textrm{and}~4$. Note that refractive
indices in regions I and IV are $\fn_1 = \fn_4 = 1$, whereas $\fn_2$ and $\fn_3$ are the refractive
indices of respectively gain and loss sections.  Thus, Helmholtz equation in (\ref{equation4}) gives
rise to $\mathcal{E}^{j}$ as follows
    \be
   \mathcal{E}^{j} (z):=a_j\,e^{i k_z \tilde{\fn}_j z} + b_j\,e^{-i k_z\tilde{\fn}_j z} ~~~~~~
{\rm for} ~~ z\in z_{j},
    \label{E-theta}
    \ee
where $a_j$ and $b_j$ are possibly $k$-dependent amplitudes, and
    \begin{align}
    \tilde\fn_j:=\sec\theta \sqrt{\fn^2_j -\sin^2\theta}.
    \label{tilde-parm}
    \end{align}
These amplitudes are associated to each other by virtue of standard boundary conditions. Since outer
interfaces of the slab are subjected to conductivities of 2D materials, they appear in boundary
conditions due to the surface current $\vec{\cJ}^{(\ell)}_j (z_{\star}) := \sigma^{(\ell)}_j(z_{\star})
\vec{\cE}^j(z_{\star})$, where $\sigma_j$ is the conductivity,  $z_{\star} = 0,~2L$ are the points
where 2D materials are placed, and $\ell$ denotes the 2D material type,
\be
\ell :=\left\{
                \begin{array}{ll}
                  g, & \hbox{for Graphene;} \\
                  w, & \hbox{for 2D WSM.}
                \end{array}
              \right. \nonumber
\ee
Apparently, boundary conditions relate coefficients $a_j$ and $b_j$, and give rise to the construction
of transfer matrix which is a useful tool to reveal lasing threshold and CPA conditions. Hence, one
incorporates right outgoing waves to the left ones by means of the transfer matrix which is defined as
     \begin{align}
    \left[\begin{array}{c}
    a_4\\ b_4\end{array}\right]=\bM \left[\begin{array}{c}
    a_1\\ b_1\end{array}\right].
    \nn
    \end{align}
Thus, lasing threshold and CPA conditions, respectively, correspond to the real zeros of $M_{22}$ and
$M_{11}$ components of $\bM$. Notice that they are complex conjugate of each other and connected to each
other via $\mathcal{PT}$ symmetry because of time reversal symmetry. Lasing threshold condition has been
studied extensively in \cite{lastpaper}, and in our context we use $M_{11} = 0$ to unveil the CPA
conditions at which $M_{11}$ is obtained explicitly as
    \be
    M_{11} =\frac{e^{-2i k_z L}}{8\tilde{\fn}_2\tilde{\fn}_3}
\left[ V_{+}\fu_{+}^{(3)}e^{ik_z L \tilde{\fn}_3} - V_{-}\fu_{-}^{(3)} e^{-ik_z L \tilde{\fn}_3} \right],
    \label{M11=x}
    \ee
where we identify
    \begin{align}
    &\fu_{\pm}^{(j)} :=  \sigma_j^{(\ell)} \pm (\tilde\fn_j + 1), \label{u=}\\
    &V_{\pm} := (\tilde{\fn}_3 \pm \tilde{\fn}_2) \fu_{+}^{(2)}e^{ik_z L \tilde{\fn}_2} +
(\tilde{\fn}_3 \mp \tilde{\fn}_2) \fu_{-}^{(2)}e^{-ik_z L \tilde{\fn}_2},
    \label{defns}
    \end{align}
and $\sigma_j^{(\ell)}$ is the conductivity of the 2D material based on the exterior surface of region
$j$. For graphene, conductivity $\sigma_j^{(g)}$ is determined within the random phase approximation in
\cite{conductivity-graphene,conductivity-graphene-1,conductivity-graphene-2} as the sum of intraband
and interband contributions, i.e. $\sigma^{(g)} = \sigma^{(g)}_{intra}+\sigma^{(g)}_{inter}$ for each
$j$, where
    \begin{align}
    \sigma^{(g)}_{intra} &:= \frac{ie^2\chi}{\pi\hbar^2\left(\omega + i\Gamma\right)}\ln\left[2\cosh\left(\frac{\mu}{\chi}\right)\right], \notag\\
    \sigma^{(g)}_{inter} &:=\frac{e^2}{4\pi\hbar}\Bigl[\frac{\pi}{2} + \arctan\left(\frac{\nu_{-}}{\chi}\right) -\frac{i}{2}\ln\frac{\nu_{+}^2}{\nu_{-}^2 + \chi^2}\Bigr].
    \label{conductivitygraphene}
    \end{align}
where we define $\nu_{\pm} := \hbar\omega \pm 2\mu$ and $\chi:=2k_B T$. Here, $-e$ is the electron charge,
$\hbar$ is the reduced Planck's constant, $k_B$ is Boltzmann's constant, $T$ is the temperature, $\Gamma$
is the scattering rate of charge carriers, $\mu$ is the chemical potential, and $\hbar\omega$ is the
photon energy~\cite{naserpour}. Likewise, for 2D WSM the conductivity for each sheet is computed by using
Kubo formalism \cite{weyl-conductivity} as
   \begin{align}
   \sigma^{(w)} \approx i \int_{\xi < \lambda} \frac{dk_z}{2\pi} \sigma^{2D} (k_z) \xi (k_z) = \frac{i e^2}{\pi h} \ln (2b\lambda),  \label{conductivityweyl}
   \end{align}
where surface state labeled by $k_z$ is localized near the sheets with a localization length
$\xi (k_z) = 2b/(b^2 -k_z^2)$, $\sigma^{2D} (k_z)$ is 2D quantized Hall conductivity,
$\sigma^{2D} (k_z) = e^2/h$, and $b$ is the measure of separation between Weyl nodes given by
$b=\textbf{b} /\hat{e_z}$ in the $k_z$-space. Here the symbol '$\approx$' is used to imply that real
part of $\sigma^{(w)}$ is negligibly small compared to the imaginary part. It is evident that both
$\sigma^{(g)}$ and $\sigma^{(w)}$ are complex-valued, and hence $\mathcal{PT}$-symmetry fulfils the
following relations
    \begin{align}
    &\fn_2\stackrel{\mathcal{PT}}{\longleftrightarrow}\fn_3,
     &&\tilde{\fn}_2\stackrel{ \mathcal{PT} }{\longleftrightarrow}\tilde{\fn}_3,
     &&\sigma_2^{(\ell)}\stackrel{ \mathcal{PT} }{\longleftrightarrow}-\sigma_3^{(\ell)}.
    \label{pt-symmetry-rels}
    \end{align}

Thus, we obtain the intriguing result that currents on the left and right sheets of 2D materials flow in
opposite directions. We emphasize that condition for CPA\footnote{Likewise, lasing threshold condition
could be obtained via $M_{22} = 0$ for real $k$ values, such that it is the complex conjugate of
(\ref{spec-sing})} is attained by the presence of self-dual spectral singularities \cite{jpa-2012} which
give rise to the real values of the wavenumber $k$ such that $M_{11} = 0$. Therefore, explicit form of
the self-dual spectral singularity condition is found by means of (\ref{M11=x}) as
    \be
    e^{2ik_z L \tilde{\fn}_3} = \frac{V_{-} \fu_{-}^{(3)}}{V_{+} \fu_{+}^{(3)}}.
    \label{spec-sing}
    \ee
Notice that effect of the 2D material is involved in quantities $\fu_{\pm}^{(2,3)}$ in (\ref{u=}).
Removing 2D materials by setting $\sigma^{(\ell)}_{j}= 0$ provides the CPA condition given in
\cite{lastpaper}.

\section{CPA Parameters, $\mathcal{PT}$-Symmetry and the Role of 2D Materials}\label{S4}

\begin{figure}
    \begin{center}
    \includegraphics[scale=.70]{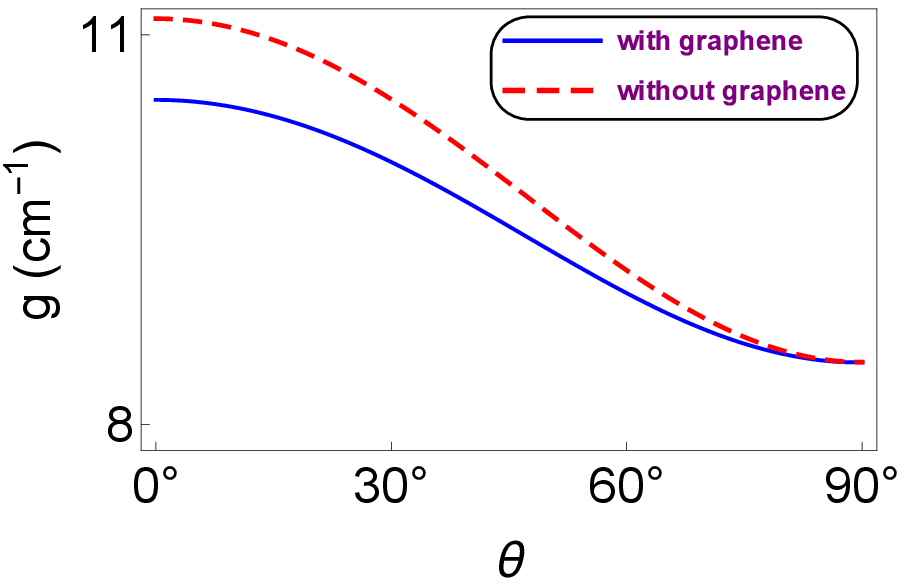}~~~
    \includegraphics[scale=.70]{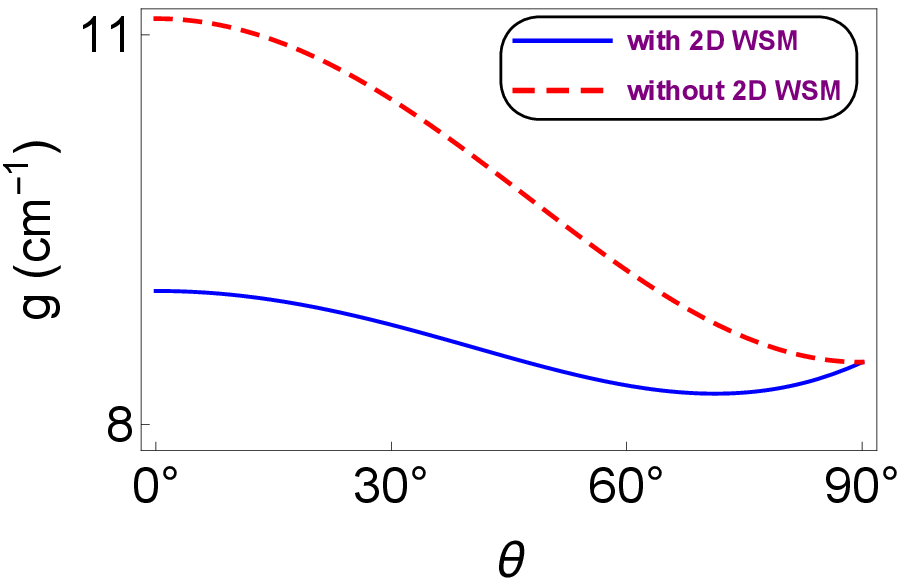}
    \caption{(Color online) Behaviour of gain amount $g$ as a function of incidence angle $\theta$
corresponding to the with and without 2D materials cases.}
    \label{fig1m}
    \end{center}
    \end{figure}

\begin{figure}
    \begin{center}
    \includegraphics[scale=.70]{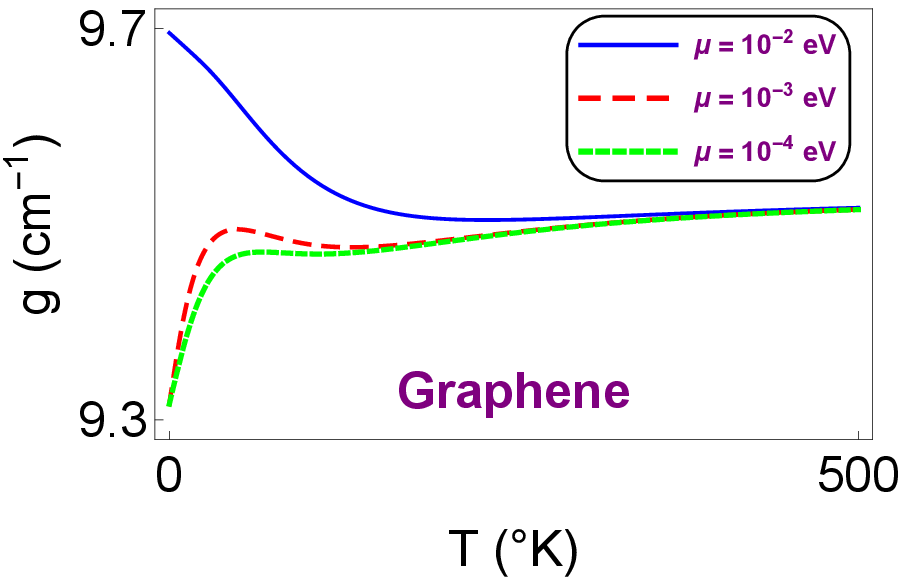}~~~
    \includegraphics[scale=.70]{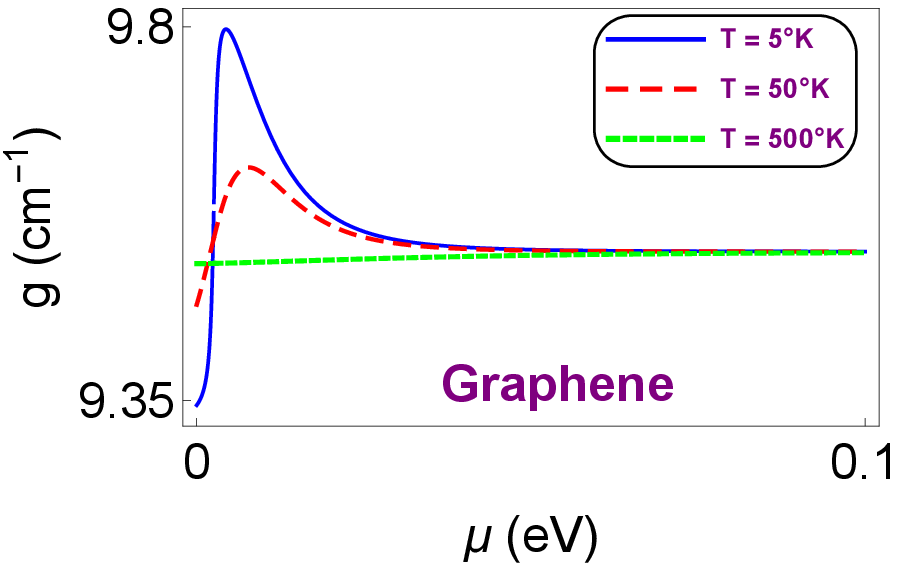}\\
    \includegraphics[scale=.70]{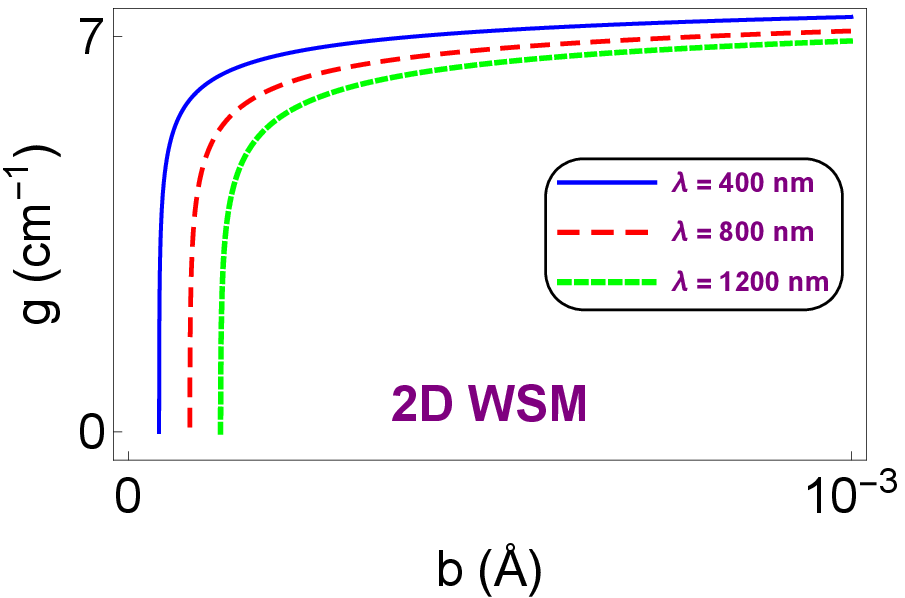}~~~
    \includegraphics[scale=.70]{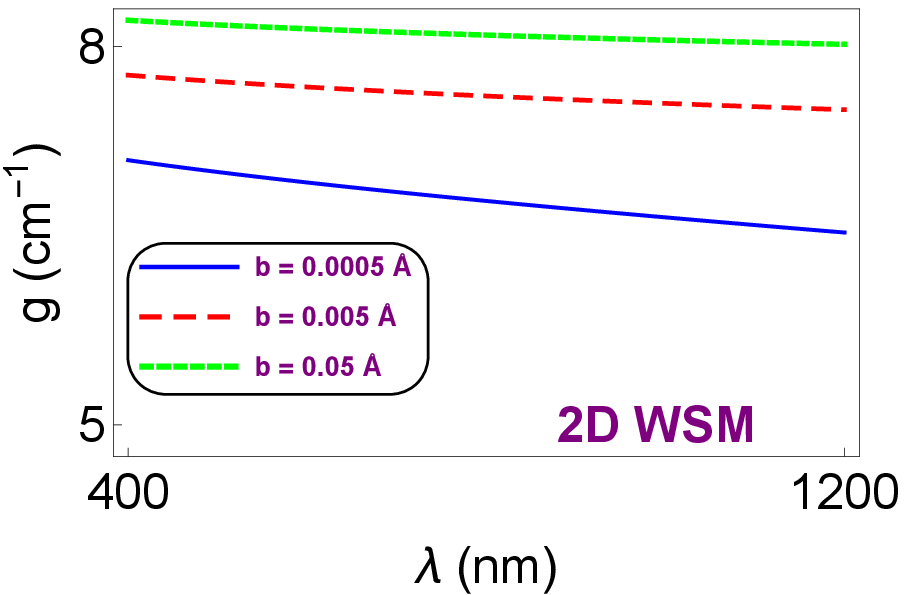}
    \caption{(Color online) Dependence of gain value on the temperature and chemical potential of
graphene sheets (top row) for graphene, and parameter $b$ and wavelength $\lambda$ for 2D WSM
(bottom row).}
    \label{fig2m}
    \end{center}
    \end{figure}

We notice that CPA condition in (\ref{spec-sing}) is the time reversal case of spectral singularities,
and thus corresponds to the spectral singularities as well for the lasing threshold condition. It is in
fact a complex expression displaying the behavior of system parameters leading to CPA conditions.Hence,
direct physical consequences of (\ref{spec-sing}) can be explored by a detailed investigation. With the
help of $\mathcal{PT}$-symmetry relations in (\ref{pt-symmetry-rels}), we reveal the following
    \begin{align}
    &\fn:= \fn_2 =\fn_3^*,
    &&\tilde\fn:= \tilde\fn_2 =\tilde\fn_3^*,
     &&\sigma^{(\ell)}:= \sigma^{(\ell)}_{2}= -\sigma^{(\ell)\ast}_{3}.
    \label{eq251}
   \end{align}
We next denote real and imaginary parts of the refractive index $\fn$ and $\tilde\fn$ as follows
    \begin{align}
    &\fn = \eta + i \kappa,
     &&\tilde\fn = \tilde\eta + i \tilde\kappa.
    \label{eq252}
    \end{align}
such that $\tilde\eta$ and $\tilde\kappa$ are written down explicitly in leading order of $\kappa$ as
    \begin{align}
    &\tilde\eta \approx \sec\theta\sqrt{\eta^2-\sin^2\theta},
    && \tilde\kappa \approx \frac{\sec\theta\eta\kappa}{\sqrt{\eta^2-\sin^2\theta}}.
    \label{eq253}
    \end{align}
This is a natural consequence of the condition $\left|\kappa\right| \ll \eta - 1 < \eta$ that most of
the materials satisfy. Afterwards, we introduce the gain coefficient $g$ as
   \begin{align}
   g:=-2k\kappa = -\frac{4\pi\kappa}{\lambda}.\label{gaincoef}
   \end{align}
We substitute (\ref{eq251}), (\ref{eq252}), (\ref{eq253}), and (\ref{gaincoef}) in the self-dual
spectral singularity condition~(\ref{spec-sing}) to obtain the lasing behavior of our system. Thus, the
most appropriate parameters of the system can be distinguished for the emergence of optimal effects. The
effect of 2D materials comes along with the conductivity $\sigma^{(\ell)}$. We elaborate an detailed
analysis towards understanding of the roles of parameters through plots of gain coefficient $g$ given in
(\ref{gaincoef}). For the slab material, we exploit Nd:YAG crystals with specifications $\eta = 1.8217$,
$L=1$~cm, and $\theta = 45^{\circ}$ for the whole system whereas we use $\Gamma = 0.1$~meV for graphene.
The remaining parameters are displayed in Figs.~\ref{fig1m} and \ref{fig2m}.

In Fig.~\ref{fig1m}, one observes impacts of the 2D materials on the threshold gain value depending
upon the incidence angle $\theta$. It is manifest that presence of 2D material reduces the required
gain amount significantly. Moreover, 2D WSM is more favorable than the graphene at moderate incidence
angles. At large angles close to the right one, effect of the 2D material is unessential. In particular,
note that 2D WSM leads to the smallest gain value even lower than the one at $\theta = 90^{\circ}$.

Fig.~\ref{fig2m} displays how properties of 2D materials influence the gain decrement. It is immediate
that adjusting parameters of 2D WSM causes the gain amount to lower considerably, even the smallest
possible gain value, compared to the adjustment of graphene parameters. As for the graphene (top row),
the maximum gain reduction happens at smaller temperatures and chemical potentials such that $\mu$ is
less than resonance frequency at which $\mu_{res} = \hbar \omega/2$ occurs. Meanwhile, as regards to
2D WSM, the extreme gain reduction is typically obtained by decreasing the wavelength $\lambda$ and
parameter $b$. It is also intriguing to observe that there is a minimum limit for the $b$ value
corresponding to each wavelength, for instance, $b_{min} \approx 0.000086$ {\AA} for $\lambda=800$~nm.
In particular, $b$ values slightly greater than $b_{min}$ yield the least gain amounts, and are
preferred for a better lasing effect, see \cite{nuh} for the possible $b$ values of a realistic 2D
WSM. Lastly, we remark that temperature (and also chemical potential) dependence of refractive indices
in both cases (in particular with the graphene case) is ignored safely since it yields negligible
effect (about 0.001\%) within the temperature ranges of interest~\cite{temp, temp2}.

\section{Broadband CPA-Laser Action with Equal Amplitude and Phase of Incoming Waves}

CPA action emerges in principle when time-reversed system satisfies the self-dual spectral singularity
condition given in (\ref{spec-sing}) simultaneously with lasing incidence. This intriguing case is a
special occasion that exists only if the correct and exact amplitudes and phases of the incoming waves
are emitted from both sides of the optical system such that overlapping waves are eventually supposed
to interfere destructively. In the case of CPA the waves outside are originated by an incidence angle
of $-\theta$, and amplitudes of incoming waves are related to each other in terms of amplitudes given
in (\ref{E-theta}) by the ratio \cite{lastpaper}
\be
\rho = \frac{a_4^{\ast}\,e^{-2ik_z L}}{b_1^{\ast}}, \label{rho=}
\ee
such that incoming waves are absorbed perfectly to build full destructive interference. In fact, $a_4$
can be expressed in terms of $b_1$ by using the spectral singularity condition. Spectral singularity
condition entails $a_1 = b_4 = 0$ such that purely outgoing waves take place. Thus, together with the
boundary conditions and CPA condition in (\ref{spec-sing}) one gets
\be
a_4 =\frac{e^{-2ik_z L}\,b_1}{\tilde{\fn}^{\ast}} \sqrt{\frac{V_{+}\,V_{-}^{\ast}}{\fu_{+}^{(2)\ast}
\fu_{-}^{(3)\ast}}}. \label{a4=}
\ee
Hence, we obtain that incoming waves are perfectly absorbed provided that $\rho$ satisfies
\be
\rho = \frac{1}{\tilde{\fn}}\sqrt{\frac{V_{+}^{\ast}\,V_{-}}{ \fu_{+}^{(2)}\fu_{-}^{(3)}}}. \label{rho2=}
\ee
Notice that one must hold incoming waves with the ratio of amplitudes $\left|\rho\right|$ and
corresponding phase difference $\delta\phi$ defined by $e^{i\delta\phi} = \rho / \left|\rho\right|$ to
obtain a CPA-laser. Although the attainment of $\left|\rho\right|$ and $\delta\phi$ yields a perfect CPA
case, the easiest condition could be adopted by setting $\rho=1$ so that amplitudes and phases of the
waves emergent outside the slab are the same. This yields
\be
V_{+}^{\ast}\,V_{-} = \tilde{\fn}^2\,\fu_{+}^{(2)}\,\fu_{-}^{(3)}\label{equalamp}
\ee
This is a complex equation that gives the parameters of a CPA yielding equal amplitudes and phases of
the incoming waves. An experimental realization of this fact in the case of dispersion could be
expressed for both graphene and 2D WSM in Table~\ref{table3}, where the slab thickness is taken as
$L=1$~cm, see appendix for the dispersion effect.
 \begin{table*}[ht]
\centering
\resizebox{\columnwidth}{!}{%
\begin{tabular}{@{\extracolsep{4pt}}llcccccccc}
\hline
{} & {} & {} & \multicolumn{3}{c}{with Graphene ($T = 300~^{\circ}K$ and $\mu = 5~\textrm{meV}$)}  & {$-$} &\multicolumn{3}{c}{with 2D WSM ( $b = 0.05~${\AA})}\\
 \cline{2-2}
 \cline{3-5}
 \cline{6-8}
   \hline
  & $\theta$ & {} & $0^{\circ}$ & $-40^{\circ}$ & $-80^{\circ}$ & {} & $0^{\circ}$ & $-40^{\circ}$ & $-80^{\circ}$ \\
  & $\lambda$ & {} & $807.996~\textrm{nm}$ & $808.009~\textrm{nm}$ & $807.994~\textrm{nm}$ & {} & $808.006~\textrm{nm}$ & $808.001~\textrm{nm}$ & $807.999~\textrm{nm}$ \\
  & $g$ & {} & $9.722~\textrm{cm}^{-1}$ & $9.087~\textrm{cm}^{-1}$ & $8.626~\textrm{cm}^{-1}$ & {} & $7.798~\textrm{cm}^{-1}$ & $8.547~\textrm{cm}^{-1}$ & $0.779~\textrm{cm}^{-1}$ \\
  & $\kappa$ & {} & $-6.251\times 10^{-5}$ & $-5.843\times 10^{-5}$ & $-5.546\times 10^{-5}$ & {} & $-5.014\times 10^{-5}$ & $-5.496\times 10^{-5}$ & $-5.011\times 10^{-6}$ \\
 \hline
\end{tabular}%
}
\caption{Physical parameters for the construction of a bidirectionally equal amplitude and phase CPA for
various incident angles corresponding to the graphene and 2D WSM cases.}\label{table3}
\end{table*}

\begin{figure}
    \begin{center}
    \includegraphics[scale=.50]{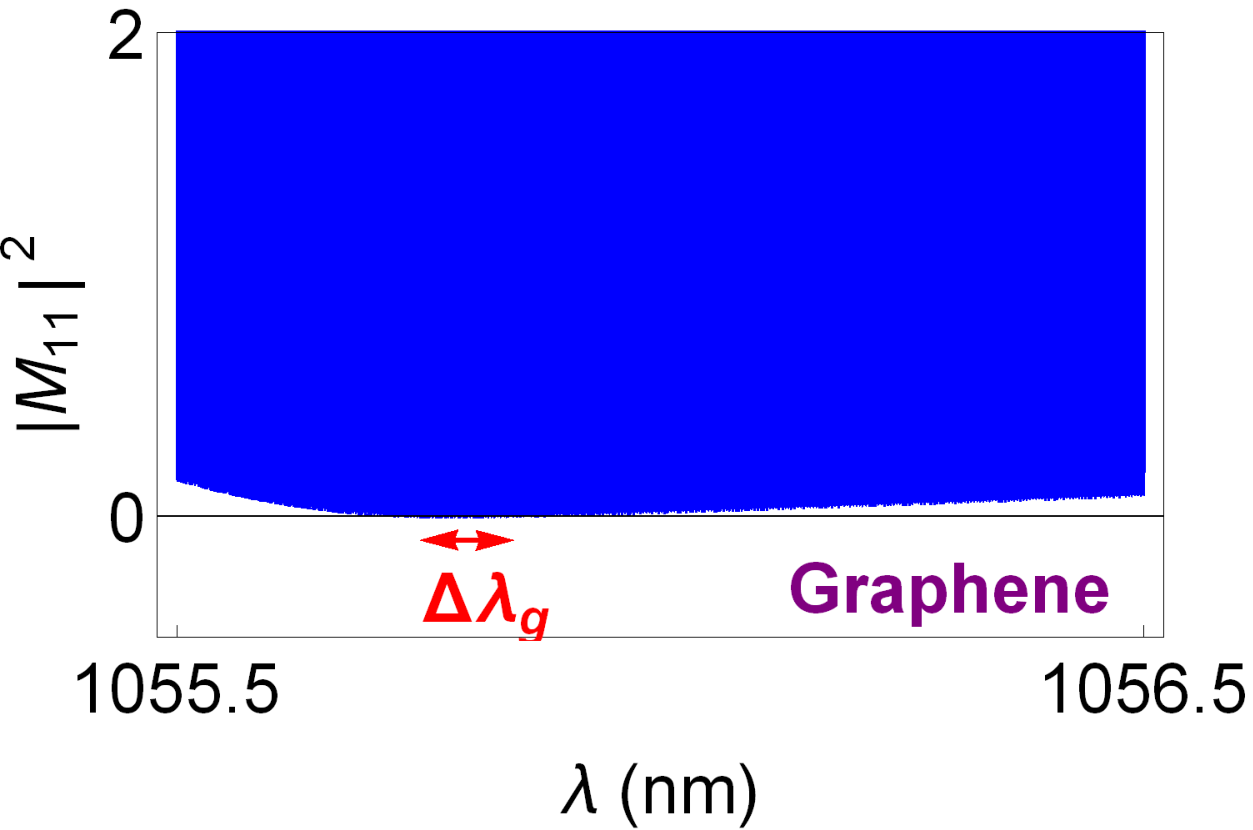}~~~
    \includegraphics[scale=.50]{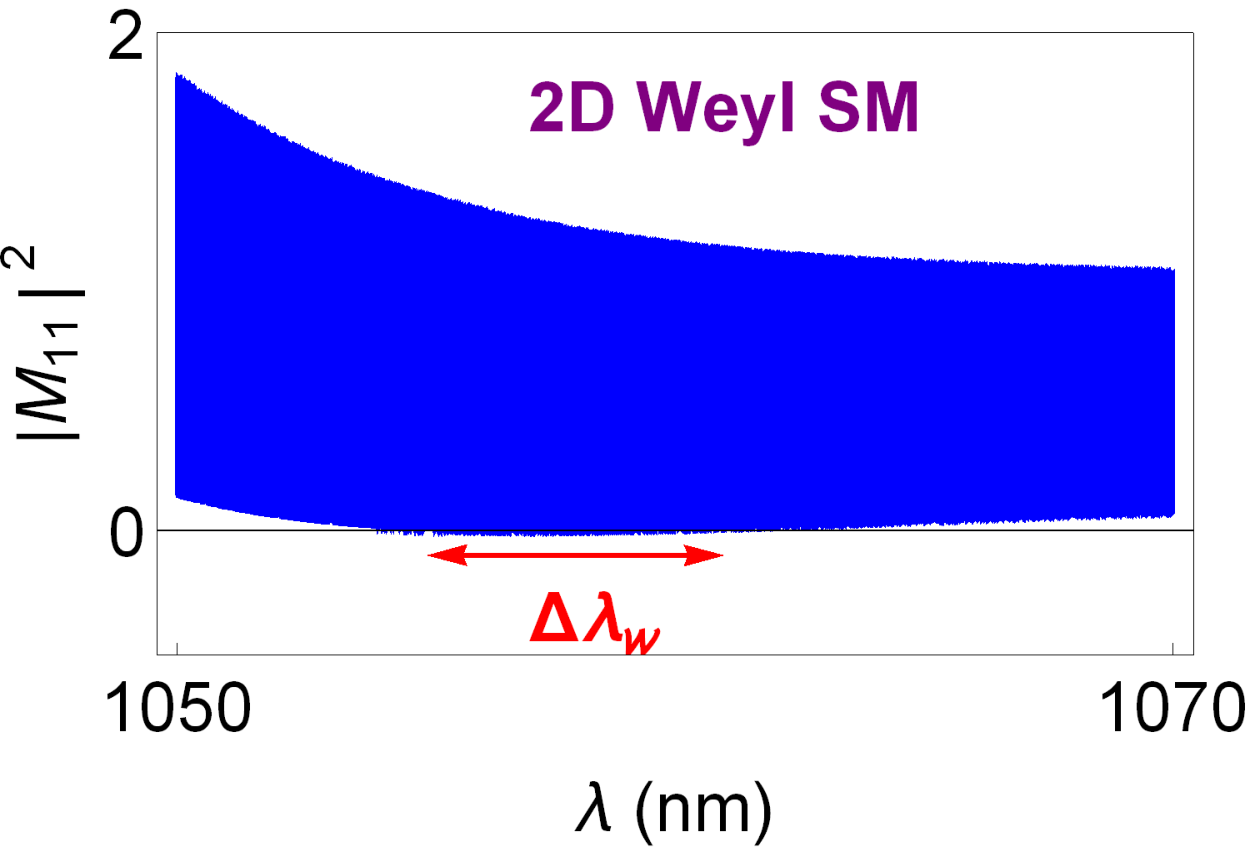}\\
    \caption{(Color online)  Broadband CPA for the case of graphene and 2D WSM.
$\Delta \lambda_g = 0.1$~nm corresponding to the range (1055.75~nm, 1055.85~nm) and
$\Delta \lambda_w = 6$~nm corresponding to the range (1055~nm, 1061~nm) give rise to the broadband CPA
within $1\%$ of flexibility. }
    \label{broadband}
    \end{center}
    \end{figure}

However, for the construction of a broadband CPA, we split the real and imaginary parts of
(\ref{equalamp}) to obtain the following
\begin{align}
&\mathfrak{a}_{-}\,\cos\left(2k_z L \tilde{\eta}\right)-2\,\textrm{Im}[\fu_{+}^{(2)}]\,\sin\left(2k_z L \tilde{\eta}\right) = \mathfrak{c}_{1}\label{equal_amp1}\\
&2\,\textrm{Im}[\fu_{+}^{(2)}]\,\cos\left(2k_z L \tilde{\eta}\right) + \mathfrak{a}_{-}\,\sin\left(2k_z L \tilde{\eta}\right) + \frac{\tilde{\kappa}}{\tilde{\eta}} \Bigl\{\mathfrak{b}_{-}\,e^{\tilde{g}L} - \mathfrak{b}_{+}\,e^{-\tilde{g}L}\Bigr\} = \mathfrak{c}_{2}\label{equal_amp2}
\end{align}
where we identified
\begin{align}
&\mathfrak{a}_{\pm} := \left|\sigma^{(\ell)} + \tilde{\fn}\right|^2 \pm 1,~~~\mathfrak{b}_{\pm} := \mathfrak{a}_{+} \pm 2\left[\tilde{\eta} + \textrm{Re}[\sigma^{(\ell)}]\right],~~~\tilde{g} := \frac{\eta\,g}{\sqrt{\eta^2 - \sin^2\theta}}\notag\\
&\mathfrak{c}_{1} :=\frac{1}{4}\left\{\tilde{\eta}^2 -1 -\left|\sigma^{(\ell)}\right|^2 - 2\,\textrm{Re}[\sigma^{(\ell)}] -4\tilde{\kappa}\,\textrm{Im}[\sigma^{(\ell)}] \right\}\nonumber\\
&\mathfrak{c}_{2} := \frac{1}{2}\left\{\tilde{\kappa}(1-\tilde{\eta}) -\tilde{\eta}\,\textrm{Im}[\sigma^{(\ell)}] +\frac{\tilde{\kappa}}{\tilde{\eta}}\left[1 + \left|\sigma^{(\ell)}\right|^2 + (2+\tilde{\eta})\,\textrm{Re}[\sigma^{(\ell)}] \right]\right\}\nonumber
\end{align}
Notice that the first equation (\ref{equal_amp1}) leads to an expression for $k$
\begin{align}
&k = \frac{m\pi}{\tilde{\eta} L \cos\theta} - \frac{1}{2\tilde{\eta} L \cos\theta} \sin^{-1}\left\{ \frac{2\,\textrm{Im}[\fu_{+}^{(2)}]\mathfrak{c}_{1} - \mathfrak{a}_{-}\sqrt{\mathfrak{a}_{-}^2 + 4 \textrm{Im}[\fu_{+}^{(2)}]^2 - \mathfrak{c}_{1}^2}}{\left[\mathfrak{a}_{-}^2 + 4 \textrm{Im}[\fu_{+}^{(2)}]^2\right]}\right\}\label{wavelength2}
\end{align}
where $m$ is the mode number that corresponds to the CPA points obtained by means of self-dual spectral
singularity condition in (\ref{spec-sing}). We observe that in fact when all other parameters are kept
fixed, the wavelength $\lambda$ is inversely proportional to the mode number $m$ according to
(\ref{wavelength2}). Thus, for each $m$, the wavelength becomes unique. If the mode number is large
enough so that self-dual spectral singularity points seem to behave as continuous, one can speak of
broadband CPA phenomenon. Hence, it is easy to show that broadband range of a CPA realization is
computed as
\be
\Delta \lambda = \frac{\lambda^2 \left|\Delta m\right|}{2 L \tilde{\eta} \cos\theta}, \label{broadband2}
\ee
where $\lambda$ corresponds to wavelength at the center of interval $\Delta \lambda$ which could be
computed via (\ref{wavelength2}). Fig.~\ref{broadband} serves as the pictorial demonstration of this
situation, which is associated with graphene in the left board and 2D WSM in the right board. For the
purpose of realizing the broadband CPA, we employ a wide slab thickness of 10~cm in size with the wave
emergent by angle $\theta = -30^{\circ}$. In the case of graphene, provided that self-dual spectral
singularity condition gives rise to the mode number $m = 331106$ corresponding to the wavelength
$\lambda = 1056.992$~nm, and the gain amount $g \approx 3734~\textrm{cm}^{-1}$ with the graphene
features of $T = 300^{\circ}\textrm{K}$ and $\mu = 0.005~\textrm{eV}$ for the chemical potential, one
obtains broadband CPA with $\Delta \lambda_g = 0.1$~nm corresponding to the wavelength range
(1055.75~nm, 1055.85~nm). Likewise, for the 2D WSM, one obtains $m = 331107$, $\lambda = 1057.145$~nm,
$g \approx 3211~\textrm{cm}^{-1}$, and parameter $b = 0.05$ {\AA} to obtain $\Delta \lambda_w = 6$~nm
corresponding to the wavelength range (1055~nm, 1061~nm) within $1\%$ of precision. This shows that
the expression (\ref{broadband2}) yields quite well estimation of broadband CPA condition once the
self-dual spectral singularity points are determined corresponding to a mode number $m$. Also, the
use of 2D WSM is favorable for the realization of broadband CPA due to its nice features present in
its conductivity expression. We stress out that this broadband structure perfectly fits an intriguing
situation of equal amplitude and phase conditions of the particular CPA provided. Finally, we realize
that Eq.~\ref{equal_amp2} enables to compute the necessary gain amount corresponding to equal amplitude
and phase conditions, which verifies the results obtained in Figs.~\ref{fig1m} and \ref{fig2m}.

\section{Concluding Remarks}
\label{S9}

In this study, we provide necessary conditions to obtain equal wave amplitudes and phases corresponding
to a $\mathcal{PT}$-symmetric CPA system which is covered by 2D materials. We impose an overall
$\mathcal{PT}$ symmetry in the system just because of the symmetry conditions arising from duality
between the lasing and CPA phenomena. We derived exact expression for the CPA condition based on the
particular 2D materials in (\ref{spec-sing}). Although our expression in (\ref{spec-sing}) holds for all
2D materials with scalar-valued conductivities, we in particular employed graphene and 2D WSM in our
analysis because of their prominent roles in recent 2D materials research.

We observe that 2D WSM is much more effective than graphene in the sense of gain decrement. In
particular, adjusting minimum $b$ value in 2D WSM gives rise to an incredibly low gain values.
Nevertheless, In the case of graphene the gain reduction remains limited. Thus, 2D WSM is more
appropriate for the construction of a good CPA. We find out that graphene is more efficient at lower
temperatures and chemical potentials below the resonance conditions. However, 2D WSM is much operative
around critical $b$ values at which it yields zero gain value. As the wavelength increases, efficiency
of CPA-laser corresponding to 2D WSM improves.

Finally, we explicitly demonstrate necessary conditions to realize a CPA which will be formed by equal
amplitudes and phases of the waves emergent from both sides. This is important because the main problem
in experimental realization of a CPA is just to adjust correct amplitudes and phases of incoming waves.
But our model rigorously resolves this complication. If the CPA parameters are tuned in according to
(\ref{equalamp}), then any equal wave amplitudes and phases fulfils a CPA-laser. We present some exact
parameters in Table~\ref{table3} to guide the experimental attempts for the realization of the
$\mathcal{PT}$-symmetric CPA phenomenon with 2D materials. We also present conditions for the
realization of a broadband CPA in (\ref{broadband2}) which occurs at large mode numbers $m$. For this
to happen, one should pick a wider slab thickness $L$ such that corresponding self-dual spectral
singularity points give rise to relatively high $m$ values. Although our model is focused on two
representative 2D material types, it comprehends set of all 2D materials whose conductivities should be
given with corresponding parameters. Novelty of our results awaits experimental verification which is
significant to establish a realistic CPA, and to address the direction of research accordingly.  \\[6pt]

\appendix
\section{Dispersion Effect}

If there exists a dispersion in the refractive index $\fn$, then we need to incorporate the effect of
wavenumber $k$ on $\fn$. We imagine that active part of the optical system composing the gain
ingredient is formed by doping a host medium of refractive index $n_0$, and its refractive index
satisfies the dispersion relation
    \be
    \fn^2= n_0^2-
    \frac{\hat\omega_p^2}{\hat\omega^2-1+i\hat\gamma\,\hat\omega},
    \label{epsilon}
    \ee
where $\hat\omega:=\omega/\omega_0$, $\hat\gamma:=\gamma/\omega_0$, $\hat\omega_p:=\omega_p/\omega_0$,
$\omega_0$ is the resonance frequency, $\gamma$ is the damping coefficient, and $\omega_p$ is the plasma
frequency. The $\hat\omega_p^2$ can be described in leading order of the imaginary part $\kappa_0$ of
$\fn$ at the resonance wavelength $\lambda_0:=2\pi c/\omega_0$ by the expression
$\hat\omega_p^2=2n_0\hat\gamma\kappa_0$, where quadratic and higher order terms in $\kappa_0$ are
ignored~\cite{pra-2011a}. We replace this equation in (\ref{epsilon}), employ the first expression of
(\ref{eq252}) and neglecting quadratic and higher order terms in $\kappa_0$, we obtain the real and
imaginary parts of refractive index as follows \cite{CPA, CPA-1, CPA-2, CPA-3, pra-2011a}
     \begin{align}
    \eta\approx n_0+\frac{\kappa_0\hat\gamma(1-\hat\omega^2)}{(1-\hat\omega^2)^2+
    \hat\gamma^2\hat\omega^2},
    \qquad \kappa\approx\frac{\kappa_0\hat\gamma^2\hat\omega}{(1-\hat\omega^2)^2+
    \hat\gamma^2\hat\omega^2}.
    \label{eqz301}
    \end{align}
$\kappa_0$ can be written as $\kappa_0=-\lambda_{0}g_0/4\pi$ at resonance wavelength $\lambda_0$, see
(\ref{gaincoef}). Substituting this relation in (\ref{eqz301}) and making use of (\ref{eq252}) and
(\ref{equalamp}), we can determine $\lambda$ and $g_{0}$ values for the CPA with equal amplitudes and
phases yielding a perfect destructive interference. These are explicitly calculated for various
parameters in Table~\ref{table3} for our setup of the $\cP\cT$-symmetric bilayer covered by 2D
materials. Furthermore, Nd:YAG crystals forming the slab material hold the following $\hat\gamma$ value
corresponding to the related refractive index and resonance wavelength  \cite{silfvast}:
    \begin{align}
    n_0=1.8217, \qquad \lambda_0=808\,{\rm nm}, \hat\gamma=0.003094.
    \label{specifications}
    \end{align}

\end{document}